\documentclass[preprint,showpacs,preprintnumbers,amsmath,amssymb]{revtex4}
\usepackage{graphicx}
\usepackage{dcolumn}
\usepackage{bm}

\newtheorem{thm}{Theorem}[section]

\newtheorem{cor}[thm]{Corollary}
\newtheorem{prop}[thm]{Proposition}

\newcommand{\C}{\mathbb{C}}

\renewcommand{\phi}{\varphi}
\renewcommand{\epsilon}{\varepsilon}
\newcommand{\M}{\mathcal{M}}

\newcommand{\R}{\mathbb{R}}

\newcommand{\tr}{{\rm tr}}

\begin{document}
\title[Unital quantum operations] {Unital quantum operations on the Bloch ball and Bloch
region}
\thanks{Research supported in part by a grant from the National Science Foundation
(DMS 0100290).}
\author{P.~S.~Bourdon}
 \affiliation{Department of Mathematics,\\ Washington and Lee University, Lexington, VA 24450} 
\email{pbourdon@wlu.edu}
\author{H.~T.~Williams}%
\affiliation{Department of Physics and Engineering,\\ Washington and Lee University,\\ Lexington, VA 24450}%
 \email{williamsh@wlu.edu}

\date{\today}

\begin{abstract}
 For one qubit systems, we present a short, elementary argument characterizing unital
quantum operators in terms of their action on Bloch vectors.  We then show how our approach generalizes
to multi-qubit systems, obtaining inequalities that govern when  a ``diagonal'' superoperator on
the Bloch region is a quantum operator. These inequalities are the $n$-qubit analogue of the
Algoet-Fujiwara conditions.   Our work is facilitated by an analysis of operator-sum decompositions in
which negative summands are allowed. 
\end{abstract}

\maketitle

\section{Introduction}  A {\it quantum operator} (or quantum superoperator) $\Phi$ on the collection
$\M_N$ of complex
$N\times N$ matrices  is a completely positive, trace preserving linear map.   The quantum operator
$\Phi$ is {\it unital} provided that $\Phi(I) = I$, that is, provided that $\Phi$ fixes the identity
matrix.  A density matrix, which represents the state of a quantum system, is a positive matrix (Hermitian
with nonnegative eigenvalues) having trace one.  The properties of density matrices are thus preserved via the action
of a quantum operation. Of course, density matrices are mapped to density matrices under any
trace-preserving positive superoperator. (A positive superoperator, by definition, takes positive
matrices to positive matrices.) The requirement that a quantum operation be completely positive
rather than  simply positive is based on the viewpoint that
$\Phi$ represents the ``restriction'' of a positive operator on a larger system.   By definition, a
superoperator $\Phi$ on $\M_N$ is completely positive provided that $I \otimes \Phi: \M_m\otimes
\M_N\rightarrow \M_m\otimes \M_N$ is positive for all positive  integers $m$.  

    A Density matrix $\rho\in \M_2$ represents the state of a two-level quantum system---a one qubit
system.  It's not difficult to show that such matrices have the following ``Bloch'' representation:
\begin{equation}\label{BVR1}
\rho = \frac{I + \sum_{i=1}^3 r_i\sigma_i}{2},
\end{equation}
where $(r_1, r_2, r_3)$ belongs to  the closed unit ball of $\R^3$ and where 
$$
\sigma_1 = \left[\begin{array}{cc}0 &1 \\1&0\end{array}\right], \sigma_2 = \left[\begin{array}{cc}0 &-i
\\i&0\end{array}\right], \sigma_3 = \left[\begin{array}{cc}1 &0 \\0&-1\end{array}\right]
$$
are the usual Pauli matrices.  The vector $\vec{r} = (r_1, r_2, r_3)$ appearing in (\ref{BVR1}) is called
the {\it Bloch vector} or {\it coherence vector} of $\rho$.  The correspondence between elements of the closed
unit ball of $\R^3$ and density matrices is complete for two-level systems; that is, a trace-one matrix
$\rho \in \M_2$ is positive if and only if it has representation (\ref{BVR1}) where $|\vec{r}| \le 1$.

  A density matrix
$\rho\in
\M_N$ has a representation analogous to (\ref{BVR1}):
\begin{equation}\label{BVRN} 
\rho = \frac{1}{N}\left(I + \sqrt{\frac{N(N-1)}{2}}\  \sum_{i=1}^{N^2 -1} r_i\lambda_i\right)
\end{equation}
where now the Bloch vector $\vec{r}$ belongs to the closed unit ball of $\R^{N^2 -1}$ and
$\{\lambda_i\}_{i=1}^{N^2-1}$  consists of elements of $\M_N$ having the following
properties:
$$
\lambda_i\ \text{is self-adjoint}\ (\lambda_i^\dagger = \lambda_i),\ \tr{\lambda_i} = 0,\ \text{and} \ 
\tr(\lambda_i\lambda_j) = 2\delta_{ij}.
$$
One may, for example, take $\vec{\lambda}$ to consist of generators of  SU$(N)$ 
(see \cite{JS,BK,K}).  In our representation  (\ref{BVRN}) of density matrices, we have adopted the
normalization factor $(N(N-1)/2)^{1/2}$ found in \cite{BK}, which forces a pure-state density
matrix to have a Bloch vector $\vec{r}$ of norm $1$.  Note that 
$\{\lambda_i\}_{i=1}^{N^2-1}$ together with the identity matrix $I$ constitutes an orthogonal basis
of
$\M_N$ relative to the Hilbert-Schmidt inner product.  In contrast to the situation for two-level
systems, the collection of Bloch vectors
$\vec{r}$ from (\ref{BVRN}) that correspond to density matrices is a proper subset of the unit ball of
$\R^{N^2 -1}$, recently characterized in \cite{BK,K}.  We will be concerned with $n$-qubit
systems, which means that in (\ref{BVRN}), $N=2^n$ and each element of $\{\lambda_i\}_{i=1}^{N^2-1}$  may be
taken to be an appropriately normalized tensor product of $n$ Pauli matrices, where $\sigma_0:= I$ is included in the
``Pauli collection'' but not all factors in the product can be $\sigma_0$---see Section \ref{FSec} below for details.

Because unital quantum operators are completely positive and preserve both the trace and the identity, associated
with any such operator
$\Phi$ on
$\M_N$, there is an $(N^2-1)\times (N^2 - 1)$ real matrix $M_\Phi$ such that
$$
\Phi\left(\frac{1}{N}\left(I + c \vec{r}\cdot \vec{\lambda}\right)\right) =
\frac{1}{N}\left(I + c M_\Phi\vec{r}\cdot \vec{\lambda}\right),
$$
where $c =\sqrt{N(N-1)/{2}}$ is the normalizing constant. 
We call $M_\Phi$ the {\it Bloch matrix} of $\Phi$.   

For $N=2$, Bloch matrices $M$ that correspond to unital
quantum operators are characterized in \cite{AF} (see also \cite{KR} and \cite{RSW}) in terms of
signed singular values of
$M$: 
\begin{equation}\label{AFC}
1 - d_3 \ge |d_1 - d_2|\ \ \text{and} \ \ 1+ d_3 \ge |d_1 + d_2|,
\end{equation}
where $d_1$, $d_2$, and $d_3$ may be taken to be the singular values of $M$ if $\det M \ge 0$ and may
be taken to be the additive inverses of the singular values of $M$ if $\det M < 0$. (For information about singular
values of matrices, the reader may consult \cite{DL, NC}, e.g.)  The inequalities (\ref{AFC}) are the Algoet-Fujiwara
conditions for a Bloch matrix corresponding to a unital quantum operator on the Bloch ball \cite{AF}. In Section 4 of
this paper, we present a short and completely elementary derivation of the Algoet-Fujiwara conditions, and in Section
5 we show how our methods yield a description of ``diagonal'' quantum operations on
$n$-qubit systems, obtaining the $n$-qubit analogue of the Algoet-Fujiwara conditions (see Theorem
\ref{TBD} below).  C.~King \cite{Kg} has obtained a related description of diagonal quantum operators on
three-state systems (i.e., single qutrit systems).

In the next
section, we discuss operator-sum representations of superoperators that map Hermitian matrices to Hermitian matrices.
As is well-known \cite{dP}, any superoperator
$\Phi$ on $\M_N$ that preserves Hermiticity must have the form
$$
\Phi(\rho) = \sum_{j=1}^k  \epsilon_j A_j \rho A_j^\dagger,
$$ 
where for each $j$, $A_j$ is an $N\times N$ matrix and $\epsilon_j\in \{1, -1\}$.  We show in Proposition \ref{MR}
below that 
 if $\Phi$ is completely positive and the operator elements
$\{A_j\}$ for $\Phi$ are linearly independent, then $\epsilon_j = 1$ for
$j=1,2,
\ldots, k$. Proposition \ref{MR} is the key lemma in our work of Sections 4 and 5.

Before concluding this introduction, we should add a remark about non-unital
quantum operators.  These superoperators  correspond to affine mappings: $(I + c\vec{r}\cdot
\vec{\lambda})/N
\mapsto (I + c(M\vec{r} + \vec{t})\cdot \vec{\lambda})/N$.   
For $N=2$, affine mappings $\vec{r}\rightarrow M\vec{r} + \vec{t}$ corresponding to quantum operators are
characterized in \cite{RSW}.  

\section{Sign patterns in operator-sum decompositions}\label{SPOSD}

Let $\Phi$ be a superoperator on $\M_N$ and suppose that for some positive 
integer $k$ there exist  $N\times N$ matrices $A_1$, $A_2$, \ldots, $A_k$   
along with ``signs'' $\epsilon_j\in \{-1, 1\}$ such that
\begin{equation}\label{OSD}
\Phi(\rho) = \sum_{j=1}^k \epsilon_j A_j \rho A_j^\dagger.
\end{equation}
The expression of the right of (\ref{OSD}) is called an {\it operator-sum 
decomposition} of $\Phi$ and $\{A_j\}_{j=1}^k$ is corresponding set of 
{\it operator elements}.  Operator-sum decompositions in which $\epsilon_j =1$ for every $j$ model system-environment
interactions (\cite{SMR}; see also \cite{KRS,NC}). For this reason, operator elements are sometimes
called ``interaction operators''.

  Observe that if 
$\Phi$ has an operator-sum decomposition (\ref{OSD}), then
$\Phi$ preserves Hermiticity; that is, 
$\Phi(\rho)^\dagger = \Phi(\rho)$ whenever $\rho$ is Hermitian.  
In \cite{dP}, de Pillis shows that every superoperator on $\M_N$ that 
preserves Hermiticity has an operator-sum decomposition.  For 
example, by de Pillis's result, the transpose operator $\Phi_T$ on $\M_2$ 
defined by $\Phi_T(\rho) = \rho^T$ must have an operator-sum 
decomposition.
A simple calculation shows that one such decomposition is given by
\begin{equation}\label{TFR}
\Phi_T(\rho) = \frac{I}{\sqrt{2}}\rho \frac{I}{\sqrt{2}} + 
\frac{\sigma_1}{\sqrt{2}} \rho  \frac{\sigma_1}{\sqrt{2}} -  
\frac{\sigma_2}{\sqrt{2}}\rho  \frac{\sigma_2}{\sqrt{2}} +  
\frac{\sigma_3}{\sqrt{2}}\rho  \frac{\sigma_3}{\sqrt{2}}.
\end{equation}
Operator-sum representations are highly non-unique; for instance  $\Phi_T$ 
is also given by
$$
 \Phi_T(\rho) = \left[\begin{array}{cc} 1&0\\0&0\end{array}\right]\rho 
\left[\begin{array}{cc} 1&0\\0&0\end{array}\right] + 
\left[\begin{array}{cc} 0&0\\0&1\end{array}\right]\rho  
\left[\begin{array}{cc} 0&0\\0&1\end{array}\right] + 
\frac{\sigma_1}{\sqrt{2}} \rho  \frac{\sigma_1}{\sqrt{2}} -  
\frac{\sigma_2}{\sqrt{2}}\rho  \frac{\sigma_2}{\sqrt{2}}.
$$
     It's obvious that $\Phi_T$ is a positive superoperator, preserving both Hermiticity and
eigenvalues.  On the other hand $\Phi_T$ 
is the canonical example of a positive operator that is not completely 
positive.  Can the fact that $\Phi_T$ is not completely positive be 
deduced from the presence of the negative sign in the operator-sum decompositions 
for $\Phi_T$ displayed above?  Proposition~\ref{MR} below shows that the answer to this question 
is yes.  This is not an entirely trivial matter.  For example, the 
identity superoperator operator $\Phi_I$, which is obviously completely 
positive,  is given by
\begin{equation}\label{IR}
\Phi_I(\rho) = (\sqrt{2}I) \rho (\sqrt{2} I) - I \rho I.
\end{equation}

  As we discussed in the Introduction, quantum operations are completely 
positive.  Observe that since Hermitian matrices are differences of 
positive matrices (immediate from the spectral decomposition), positive 
(and hence completely positive) superoperators must preserve Hermiticity.  Thus any quantum operator $\Phi$ has an
operator-sum  decomposition (\ref{OSD}).   In Theorem 1 of \cite{Ch}, Choi shows that a 
completely positive operator $\Phi$ has an operator-sum decomposition 
(\ref{OSD}) in which each sign is positive ($\epsilon_j = 1$ for every 
$j$).  Of course this doesn't mean that every operator-sum decomposition 
of a completely positive map must feature only positive signs, as 
(\ref{IR}) shows.  Choi does not prove his Theorem 1 as a corollary of de 
Pillis's theorem for Hermiticity-preserving superoperators.  Rather, 
he gives an elegant independent proof that also yields de Pillis's 
characterization (as Choi points out \cite[p.~277]{Ch}).  More important 
for our purposes is that Choi investigates the relationship between 
different operator-sum representations of the same superoperator, proving 
\cite[Remark 4]{Ch} that if $\{A_j\}_{j=1}^k$ and $\{E_j\}_{j=1}^m$ are 
collections of operator elements for the same superoperator on $\M_N$ and if $\{A_j\}_{j=1}^k$
is linearly independent in $\M_N$,  
then there is an isometric $m\times k$ matrix   $[\alpha_{jn}]$ such that for each $j\in
\{1, 2, 
\ldots, m\}$
\begin{equation}\label{OER}
E_j = \sum_{n=1}^k {\alpha_{jn}} A_n.
\end{equation}
Note that because $[\alpha_{jn}]$ is an isometry, we must have $m\ge k$; if {\it both}
$\{A_j\}_{j=1}^k$ and
$\{E_j\}_{k=1}^m$ are  linearly independent collections of operator elements for the
same  superoperator, then Choi shows $m=k$ and the matrix of scalars $[\alpha_{jn}]$ 
relating them by (\ref{OER}) is unitary.  
\begin{prop}\label{MR}  Suppose that $\Phi: \M_N\rightarrow \M_N$ is given by
$$
\Phi(\rho) = \sum_{j=1}^k \epsilon_j A_j \rho A_j^\dagger
$$
where $\{A_j\}_{j=1}^k$ is linearly independent in $\M_N$ and 
$\epsilon_j\in \{-1,1\}$ for $j =1$, $2$, \ldots, $k$.  Then $\Phi$ is 
completely positive if and only if $\epsilon_j = 1$ for each $j\in \{1, 2, 
\ldots, k\}$. 
\end{prop}
Proof.  If each sign ``$\epsilon_j$'' is positive, then $\Phi$ is 
easily seen to be completely positive (and the independence of $\{A_j\}$ 
is irrelevant). 

 Conversely, suppose that $\Phi$ is completely positive and has the form 
displayed in the statement of the proposition with $\{A_j\}_{j=1}^k$ 
independent.  We assume, in order to obtain a contradiction, that some of 
the signs $\epsilon_j$ are $-1$. Without loss of generality, we take 
$\epsilon_j = -1$ for $j=1$ to $p$ for some $p\in \{1, 2, \ldots, k-1\}$.   
(Clearly, not all of the signs can be negative: the linear independence of 
the set of operator elements means that no element $A_j$ is the zero 
matrix so that if all signs were negative,  $\Phi$ would map positive 
matrices to negative ones and hence couldn't be completely positive).  

Because $\Phi$ is completely positive, Choi's work shows that there exists 
an operator-sum decomposition for $\Phi$ with all signs positive:
$$
\Phi(\rho) = \sum_{j=1}^m E_j \rho E_j^\dagger.
$$
We have for every $N\times N$ matrix $\rho$,
$$
\sum_{j=1}^k \epsilon_j A_j \rho A_j^\dagger = \sum_{j=1}^m E_j 
\rho E_j^\dagger,
$$
or 
\begin{equation}\label{TDR}
\sum_{j=p+1}^{k } A_j \rho A_j^\dagger  = \sum_{j=1}^m  E_j \rho E_j^\dagger 
+ \sum_{j=1}^{p}A_j \rho A_j^\dagger,
\end{equation}
so that we have two different operator-sum representations  for the same 
superoperator $\rho \mapsto \sum_{j=p+1}^{k } A_j \rho A_j^\dagger$.  Thus, 
in particular, there are scalars $(\alpha_n)$ (forming one row of the isometric matrix
relating the operator elements on the left of (\ref{TDR}) to those on the right) such that
$$
A_1 = \sum_{n=p+1}^k \alpha_n A_n,
$$
contradicting the linear independence of $\{A_j\}_{j=1}^k$.~~QED

In the sequel, we will use the following quick corollary of the preceding 
proposition.
\begin{cor}\label{CCPS}  Suppose that $\Phi$ is a completely positive superoperator on 
$\M_N$ having the representation
$$
\Phi(\rho) = \sum_{j=1}^k \beta_j A_j \rho A_j^\dagger,
$$
where $\{A_j\}_{j=1}^k$ is linearly independent in $\M_N$ and $\beta_j$ is 
real for $j=1$, $2$, \ldots, $k$.  Then $\beta_j \ge 0$ for $j\in \{1, 2, 
\ldots, k\}$.
\end{cor}

Because the Pauli matrices $\sigma_1$, $\sigma_2$, and 
$\sigma_3$ together with the $2\times 2$ identity matrix $\sigma_0$ form a linearly independent set in $\M_2$, the
preceding corollary  shows that the superoperator $\Phi$ defined on $\M_2$ by
\begin{equation}\label{phisig}
\Phi(\rho) = \beta_0 \sigma_0 \rho \sigma_0 + \beta_1 \sigma_1 \rho \sigma_1 + \beta_2 
\sigma_2 \rho \sigma_2 + \beta_3 \sigma_3 \rho \sigma_3
\end{equation}
is completely positive only when $\beta_j \ge 0$ for $j =0$, $1$, $2$, 
$3$.   It's not difficult to obtain a characterization of {\it positivity} 
for the superoperator $\Phi$ defined by (\ref{phisig}).  The 
characterization, presented in the next proposition, shows that if $\Phi$ 
is positive but not completely positive and $\Phi$ is written in the form of (\ref{phisig}) above, then exactly one of
the scalars 
$\beta_j$ will be negative, as illustrated in equation 
(\ref{TFR}).

\begin{prop}\label{CP} The superoperator $\Phi:\M_2\rightarrow \M_2$ 
defined by (\ref{phisig}) is positive if and only if every pair from $\{\beta_j\}_{j=0}^3$ sums to a nonnegative
number:
\begin{equation}\label{PPS}
\beta_0 + \beta_1 \ge 0, \beta_0 + \beta_2 \ge 0, \beta_0 + \beta_3 \ge 0, 
\beta_1 + \beta_2 \ge 0, \beta_1 + \beta_3 \ge 0, \beta_2 + \beta_3 \ge 0.
\end{equation}
\end{prop}
Proof. The following observation will facilitate some calculations in the proof; it will also play in a crucial
role in the  final two sections of this paper.
\begin{equation}\label{PPP}
\sigma_i \sigma_j \sigma_i = \pm \sigma_j, \ \ i, j\in \{0, 1, 2, 3\},
\end{equation}
where the sign is positive when $i = 0$, $j=0$, or $i=j$, and negative otherwise. 

 We assume that $\Phi$, 
defined by (\ref{phisig}), is positive and obtain the inequalities stated 
in the proposition. Suppose $\rho$ is a positive matrix. Without loss of 
generality we will assume that $\rho$ has trace $1$ and hence has the form
$(I + r_1\sigma_1 + r_2\sigma_2 + r_3\sigma_3)/2$, where $\vec{r} = (r_1, 
r_2, r_3)$ lies in the unit ball of $\R^3$.  A simple calculation shows 
\begin{equation}\label{ROC}
\Phi(\rho) = \frac{ s_0I + r_1s_1\sigma_1+ r_2s_2\sigma_2 + r_3s_3\sigma_3}{2},
\end{equation}
where $s_0=\beta_0 + \beta_1 + \beta_2 + \beta_3$, $s_1 = \beta_0 + \beta_1 -
\beta_2 - \beta_3$, $s_2= \beta_0 - \beta_1 + \beta_2 -
\beta_3$, and $s_3 = \beta_0 - \beta_1 - \beta_2 + \beta_3$
Because $\Phi(\rho)$ is positive, its trace is nonnegative; thus,
$$
\tr(\Phi(\rho)) = s_0  \ge 0.
$$
If $s_0=0$, then the positive matrix $\Phi(\rho)$ is the zero matrix (independent of $\rho$), which, in view
of  (\ref{ROC}), forces $s_1$, $s_2$, and $s_3$ to be zero as well.  It follows that $\beta_j
= 0$ for each $j$,  and the inequalities (\ref{PPS}) hold for this case.  

Suppose $\tr(\Phi(\rho)) >0$.  Then we may rewrite  the right-hand side of 
(\ref{ROC}) as follows:
\begin{equation}\label{wFr}
s_0\left(\frac{I + 
r_1\frac{s_1}{s_0}\sigma_1 + r_2\frac{s_2}{s_0} +
r_3\frac{s_3}{s_0}\sigma_3}{2}\right),
\end{equation}
which represents a positive matrix if and only if
\begin{equation}\label{PC}
 \left(r_1\frac{s_1}{s_0}\right)^2+ \left(r_2\frac{s_2}{s_0}\right)^2 + 
\left(r_3\frac{s_3}{s_0}\right)^2 \le 1.
\end{equation}
Substituting, respectively,  $\vec{r} = (1,0,0)$, $(0,1,0)$, and 
$(0,0,1)$ into (\ref{PC}) yields the following three inequalities
$$
|s_1| \le s_0, |s_2| \le s_0, |s_3| \le s_0, 
$$
which, in turn, yield the desired inequalities (\ref{PPS}).  

    For the proof of the converse, suppose that $\Phi$ acts on the positive matrix $(I + 
\vec{r}\cdot\vec{\sigma})/2)$ and that the inequalities  
(\ref{PPS}) hold.  Adding the first and last inequalities of (\ref{PPS}), we must have
$\beta_0 + \beta_1 +  \beta_2 + \beta_3 \ge 0$.  If $\beta_0 + \beta_1 +  
\beta_2 + \beta_3  = 0$, then by grouping summands appropriately, one 
obtains that each of the inequalities of (\ref{PPS}) must be an equality 
and it follows from (\ref{ROC}) that $\Phi$ is the zero operator.  On the 
other hand, if  $\beta_0 + \beta_1 +  \beta_2 + \beta_3 > 0$ , then it's 
easy to see that the quotients multiplying $r_1$, $r_2$, and $r_3$ in 
(\ref{wFr}) must each have absolute value less than or equal to $1$ and 
this shows that the quantity on the left of (\ref{PC}) is $\le r_1^2 + 
r_2^2 + r_3^2$, which is $\le 1$ since $\rho$ is positive.  Thus 
$\Phi(\rho)$ is positive, as desired.~~QED

\section{Unitary superoperators and rotations of the Bloch region}

  In this section, we summarize known information about quantum operators having an
operator-sum decomposition with single unitary operator element.  

Suppose that
$\Phi:
\M_N\rightarrow
\M_N$ is unitary in the sense that  it has an operator-sum representation of the form
$$
\Phi(\rho) = U \rho U^\dagger,
$$
where $U$ is  a unitary $N\times N$ matrix.    Clearly such unitary 
$\Phi$'s  are unital quantum operators.  

It's well known (see, e.g., \cite[Exercise 8.13]{NC}) that
if 
$\Phi$ is unitary and acts on $\M_2$ (the one qubit situation), then its Bloch matrix $M_\Phi$ is a
rotation matrix on 
$\R^{3}$; that is, $M_\Phi^T M_\Phi = I$ and $\det (M_\Phi) = 1$.  Furthermore, it's not difficult
to show that  the correspondence between rotation matrices on $\R^3$ and unitary 
superoperators is complete in the $N=2$ setting; that is, given any rotation matrix $A$ there 
is a unitary $2\times 2$ matrix $U$ such that 
$$
U\left(\frac{I + \vec{r}\cdot \vec{\sigma}}{2}\right)U^\dagger = 
\left(\frac{I + A\vec{r}\cdot \vec{\sigma}}{2}\right).
$$

  Now suppose $N> 2$ and $\Phi:\M_N\rightarrow \M_N$ is a unitary quantum operator.  As one
would expect, once again $\Phi$ has a Bloch matrix $M_\Phi$ that is a ``rotation'', where by
rotation we mean
$M_\Phi$ is orthogonal ($M_\Phi^T M_\Phi = I$) and orientation preserving ($\det M = 1$).  It's
very easy to see that $M_\Phi$ must be orthogonal: let $\Phi(\rho) = U \rho U^\dagger$ and
$\rho = (I + c\vec{r}\cdot\vec{\lambda})/N$ and note
\begin{align}
\frac{1}{N} + (1-\frac{1}{N})|\vec{r}|^2 = \tr(\rho^2) =  \tr (U \rho U^\dagger U \rho U^\dagger)
= \tr\left(\frac{I + c M_\Phi\vec{r}\cdot\vec{\lambda}}{N}\right)^2 = \frac{1}{N}
+ (1-\frac{1}{N})|M_\Phi\vec{r}|^2 
\end{align}
so that $|M_\Phi \vec{r}| = |\vec{r}|$.  Thus $M_\phi$ is an isometry and since it has real entries, 
$M_\Phi$ is orthogonal.  The proof that $\det(M_\Phi) = 1$, which we will also present, requires a bit more effort.  
Because
$U$ is unitary, there is an orthonormal basis $(|v_j\rangle)_{j=1}^N$ of $\C^N$ consisting of eigenvectors of $U$
with corresponding eigenvalues
$(e^{ia_j})_{j=1}^N$, where the
$a_j$'s are real.  For $s\in [0, 1]$,  define 
$$
U_s  = \sum_j e^{i(1-s)a_j} |v_j\rangle\langle v_j|,
$$
so that $U_0 = U$ and $U_1 = I$.  Let $\Phi_s:\M_N \rightarrow \M_N$ be given by $\Phi_s(\rho) =
U_s \rho U_s^\dagger$ and let $M_{\Phi_s}$ be the corresponding Bloch matrix.  We have already
shown that $M_{\Phi_s}$ is orthogonal for each $s$ in $[0,1]$.  Hence $\det(M_{\Phi_s}) = \pm 1$
for each such $s$.  Since $\det(M_{\Phi_s})$ varies continuously  with $s$ and since
$\det(M_{\Phi_1}) = \det I = 1$, we see $\det(M_\Phi) = \det(M_{\Phi_0}) = 1$, as desired.  

When
$N > 2$, the correspondence between rotations and unitary quantum  operators is complicated; for
example, the angle
$\theta$  between pure-state Bloch vectors $\vec{r_1}$ and $\vec{r_2}$  must 
satisfy $\cos(\theta)  \ge -1/(N-1)$, or, equivalently, $\vec{r_1}\cdot\vec{r_2} \ge -1/(N-1)$  \cite{JS}.

  Returning to the one-qubit situation, suppose that $M$ is an arbitrary 
$3\times 3$ real matrix and the linear superoperator $\Phi_M:\M_2 \rightarrow 
M_2$ is defined by
\begin{equation}\label{phiM}
\Phi_M\left(\frac{I + \vec{r}\cdot\vec{\sigma}}{2}\right) = \frac{I + 
M\vec{r}\cdot\vec{\sigma}}{2}.
\end{equation}
An interesting problem is to determine when $\Phi_M$ is a quantum 
operation.  An obvious necessary condition is $\|M\| \le 1$, where $\|M\| 
= \max\{|M\vec{r}|: |\vec{r}| = 1\}$.       
A complete description of those $M$ such that $\Phi_M$ is a quantum 
operator may be found in \cite{AF} (see also  \cite{KR} and \cite{RSW}).  
This description is based on the singular-value decomposition of $M$, which, in turn, easily
yields the following.
\begin{prop}\label{DC} Suppose $M$ is a $3\times 3$ matrix with real entries. 
Then there exist $3\times 3$ rotation matrices $A$ and $B$ as well as a
$3\times 3$ diagonal matrix $D$ with real entries such that
$$ 
M= B D A.
$$
Moreover, if
$\det M \ge 0$, then the diagonal entries of $D$ may be chosen to be the
singular values of $M$ listed in decreasing order; otherwise, the diagonal
entries of $D$ may be chosen to be the negatives of the singular values of
$M$  listed in increasing order.
\end{prop}

Let $\Phi_M$ be the unital
superoperator (\ref{phiM}) on $\M_2$ induced by the real $3\times 3$ matrix  $M$.  Let  
$M = BDA $ be the factorization of $M$ promised by Proposition~\ref{DC}, and let
 let
$U_A$ and $U_B$ be the unitary $2\times 2$ matrices such that 
\begin{equation}\label{UUD}
U_A\left(\frac{I + \vec{r}\cdot\vec{\sigma}}{2}\right)U_A^\dagger = \frac{I +
A\vec{r}\cdot\vec{\sigma}}{2}\ \text{and} \ U_B\left(\frac{I +
\vec{r}\cdot\vec{\sigma}}{2}\right)U_B^\dagger = \frac{I + B\vec{r}\cdot\vec{\sigma}}{2}.
\end{equation}
Finally, let $\Phi_D$ be the unital superoperator defined by 
$$
\Phi_D\left(\frac{I + \vec{r}\cdot\vec{\sigma}}{2}\right) = \frac{I +
D\vec{r}\cdot\vec{\sigma}}{2}.
$$
Note that 
\begin{equation}\label{MUU}
\Phi_M(\rho) = (\Psi\circ\Phi_D\circ\Omega)(\rho),
\end{equation}
where $\Psi(\rho) = U_B \rho U_B^\dagger$ and $\Omega (\rho) = U_A\rho U_A^\dagger$.  Because
$\Psi$ and $\Omega$  and their inverses are quantum operations and compositions of quantum
operations are quantum operations (\ref{MUU}) shows that $\Phi_M$ is a quantum operation if
and only if $\Phi_D$ is a quantum operation.    
 Thus, to characterize unital quantum operators on the Bloch ball, one need only 
understand which diagonal matrices $M$ are such that $\Phi_M$ is a quantum operator.  
Necessary and sufficient conditions on the diagonal entries of $M$ (which are given 
by (\ref{AFC}) in the Introduction) that ensure $M$ induces a quantum 
operation are obtained in \cite{AF}  as well as \cite{RSW}  and \cite{KR}.  The 
method employed in \cite{AF} and \cite{RSW}  to obtain the conditions is based on the proof of
Theorem  1 of \cite{Ch}:   one analyzes the positivity of 
$$
(I\otimes \Phi_M)(E)
$$
where $I$ is the identity on $\M_2$ and where $E$ is the $4\times 4$ matrix composed 4 elementary $2\times 2$ 
blocks:
$$
E =\left[\begin{array}{cccc}1 & 0 & 0 & 1\\0&0&0&0\\0&0&0&0\\1&0&0&1\end{array}\right].
$$
The method employed in \cite{KR} involves starting with an operator-sum decomposition of the
unital quantum operator in question and expressing its operator elements as linear combinations of
the Pauli matrices.

In the next section, we take a new approach to characterizing the 
diagonal matrices corresponding to quantum operators on the Bloch ball.  
Our approach is based on our work with sign patterns in operator-sum 
decompositions in Section \ref{SPOSD} and allows convenient 
generalization to the $n$-qubit situation. 
\section{Diagonal quantum operators on the Bloch ball}

  Suppose that 
$$
D= \left[\begin{array}{ccc}d_{11} &0&0\\0&d_{22}&0\\0&0&d_{33}\end{array}\right]
$$
and $\Phi_D$  is the linear superoperator on $\M_2$ defined by
\begin{equation}\label{PMD}
\Phi_D\left(\frac{I + \vec{r}\cdot\vec{\sigma}}{2}\right) = \frac{I + 
D\vec{r}\cdot\vec{\sigma}}{2}.
\end{equation}
Letting $\vec{r} = 0$, shows $\Phi_D(I) = I$.  Even if one assumes 
that (\ref{PMD}) holds only when $(I + \vec{r}\cdot\vec{\sigma})/{2}$ is 
positive (that is when $|\vec{r}| \le 1$), then (\ref{PMD}), combined with the 
linearity of $\Phi_D$, completely determines $\Phi_D$. Letting
$\vec{r} = (1,0,0)$, we obtain $\Phi_D((I + \sigma_1)/2) = (I + d_{11}\sigma_1)/2$.  Hence, by linearity,
\begin{align*}
\Phi_D(\sigma_1) = 2\Phi_D\left( \frac{I + \sigma_1}{2} - \frac{I}{2}\right)
 = 2\Phi_D\left(\frac{I + \sigma_1}{2}\right) - 2\Phi_D\left(\frac{I}{2}\right) = d_{11}\sigma_1 
\end{align*}
Similarly, $\Phi_D(\sigma_2) = d_{22}\sigma_2$ and $\Phi_D(\sigma_3) = d_{33}\sigma_3$.  Thus
$\Phi_D$ is a diagonal operator on $\M_2$ with respect to the basis $(\sigma_0, \sigma_1, \sigma_2,
\sigma_3)$.  Of course, $\Phi_D$ is completely determined by its action
on this basis.

Because
$$
\sigma_i \sigma_j \sigma_i = \pm  \sigma_j,
$$
the superoperator $\Psi$ on $\M_2$, defined by 
\begin{equation}\label{Eform}
\Psi(\rho) = \sum_{i=0}^3 \beta_i \sigma_i\rho \sigma_i,
\end{equation} 
for some real constants $\{\beta_j\}_{j=0}^3$,  will have
$I$, $\sigma_1$, $\sigma_2$, and $\sigma_3$ as eigenvectors. Thus $\Psi$ will equal $\Phi_D$ if we can
arrange to have
$\Psi$ yield the appropriate corresponding eigenvalues: $1$, $d_{11}$, $d_{22}$, and $d_{33}$.  This is a simple matter
of solving the following linear system, the $j$-th equation of which is obtained by substituting $\rho= \sigma_j$ into
(\ref{Eform}):
\begin{eqnarray}\label{OQS}
1& =& \beta_0 + \beta_1 + \beta_2 +\beta_3\\ \nonumber
d_{11} &=& \beta_0 + \beta_1 - \beta_2 - \beta_3\\ \nonumber
d_{22}& =& \beta_0 - \beta_1 + \beta_2 -\beta_3\\ \nonumber
d_{33}& =& \beta_0 - \beta_1 - \beta_2 + \beta_3\ .
\end{eqnarray}
Let $C$ denote the $4\times 4$ matrix of coefficients of the preceding system and observe that $C$ is a symmetric
matrix such that
$C^2 = 4I$.  This observation permits quick solution of the system:
\begin{align}\label{Bvalues}
\beta_0 = \frac{1 +d_{11} + d_{22} + d_{33}}{4}, \beta_1 = \frac{1 + d_{11} - d_{22} - d_{33}}{4},
\\ \nonumber \beta_2 =
\frac{1 - d_{11} + d_{22} - d_{33}}{4}, \beta_3 = \frac{1 - d_{11} - d_{22} + d_{33}}{4}.
\end{align}

Hence we see that the superoperator $\Phi_D$ has operator-sum decomposition given by 
$$
\Phi_D(\rho) = \sum_{i=0}^3 \beta_i \sigma_i\rho \sigma_i,
$$
with the constants $\beta_j$ given by (\ref{Bvalues}).   By Corollary~\ref{CCPS},  $\Phi_D$ is
completely positive if and only if $\beta_j\ge 0$ for $j = 0$, $1$, $2$, 	and $3$. Thus we have arrived at our desired
characterization of diagonal quantum superoperators on $\M_2$. Observe that the nonnegativity of the
$\beta_j$'s is equivalent to the Algoet-Fujiwara conditions (\ref{AFC}).  

 Combining our work on diagonal superoperators with the factorization (\ref{MUU}), we
find an operator-sum decomposition of the unital superoperator $\Phi_M$ defined by (\ref{phiM}):
\begin{equation}\label{OSDLS}
\Phi_M(\rho) = \sum_{i=0}^3 \beta_i(U_B\sigma_iU_A)\rho
(U_B\sigma_iU_A)^\dagger
\end{equation}
where $M = B D A$ is the factorization of Proposition~\ref{DC}, where $U_A$ and $U_B$ are the
unitary matrices given by (\ref{UUD}), and where the scalars $\beta_i$ are defined by (\ref{Bvalues}) .  As discussed
above the superoperator
$\Phi_M$ will be completely positive if and only if the scalars $\beta_i$ leading each summand are nonnegative. In
\cite[Theorem 1(1)]{LS}, Landau and Streater show that every unital quantum superoperator on $\M_2$ is a
convex combination of unitary maps.    Observe that (\ref{OSDLS}) recaptures the Landau-Streater
result, and says a bit more: {\it every unital superoperator $\Phi$
on $\M_2$ that preserves both Hermiticity and 
trace is a linear combination of unitary superoperators: $\Phi(\rho) = \sum_{i=0}^3 \beta_iU_i\rho
U_i^\dagger$ where $\sum_{i=0}^3 \beta_i = 1$ for real, but not necessarily positive, scalars $\beta_i$}. 
\section{Diagonal quantum operators on the Bloch region}\label{FSec}

  Let $S = \{0, 1, 2, 3\}$  be the index set for the Pauli matrices (including $\sigma_0 = I$) and
let $S_0^n = S^n\setminus \{(0,0,\ldots,0)\}$ be the Cartesian product of $n$ copies of $S$ with the
zero $n$-tuple removed. We represent the state
$\rho$ of a
$n$-qubit system in Bloch form
$$
 \frac{1}{2^n}\left(I + \sqrt{{2^{n-1}(2^n-1)}} \sum_{i=1}^{2^{2n} -1} r_i\lambda_i\right),
$$
where $\{\lambda_i\}_{i=1}^{2^{2n} -1}$  consists of all (appropriately normalized) $n$-factor
tensor products of the Pauli matrices, excluding $I = \sigma_0\otimes\cdots\otimes\sigma_0$:
$$
\{\lambda_i\}_{i=1}^{2^{2n} -1} = \{\frac{1}{\sqrt{2^{n-1}}}\sigma_{j_1}\otimes\sigma_{j_2}\otimes
\cdots \otimes\sigma_{j_n}: (j_1, j_2,
\ldots, j_n)\in S_0^n\}.
$$
Observe that $\{\lambda_i\}_{i=1}^{2^{2n} -1}$ together with $\lambda_0:= I/\sqrt{2^{n-1}}$
constitutes an orthogonal basis for $\M_{2^n}$ such that $\langle \lambda_i|\lambda_j\rangle =
2\delta_{ij}$, where $\langle \cdot| \cdot\rangle$ is the Hilbert-Schmidt inner product: $\langle
A|B\rangle = \tr(A^\dagger B)$.  

 A basis should be ordered and we will use the ``dictionary'' ordering:
\begin{align*}
\lambda_0 = \frac{1}{\sqrt{2^{n-1}}}\sigma_0\otimes\sigma_0\otimes\cdots\otimes\sigma_0\otimes
\sigma_0,\  
\lambda_1 =
\frac{1}{\sqrt{2^{n-1}}}\sigma_0\otimes\sigma_0\otimes\cdots\otimes
\sigma_0\otimes \sigma_1,\\
\lambda_2 =
\frac{1}{\sqrt{2^{n-1}}}\sigma_0\otimes\sigma_0\otimes\cdots\otimes
\sigma_0\otimes \sigma_2, \
\lambda_3 =
\frac{1}{\sqrt{2^{n-1}}}\sigma_0\otimes\sigma_0\otimes\cdots\otimes
\sigma_0\otimes \sigma_3,\\
\lambda_4 =
\frac{1}{\sqrt{2^{n-1}}}\sigma_0\otimes\cdots\otimes\sigma_0\otimes
\sigma_1\otimes \sigma_0, \ldots,
\lambda_{2^{2n} -1} =
\frac{1}{\sqrt{2^{n-1}}}\sigma_3\otimes\sigma_3\otimes\cdots\otimes
\sigma_3\otimes \sigma_3.
\end{align*}
Note that the dictionary ordering is equivalent to that produced by ordering according to the size of the index
sequence
$i_1i_2\ldots i_n$ associated with $\sigma_{i_1}\otimes\cdots\otimes\sigma_{i_n}$, where $i_1i_2\ldots
i_n$ is interpreted as the base 4 representation of a number.  

Our goal is to characterize those $(2^{2n}-1)\times (2^{2n}-1)$
diagonal matrices
$D$, with real entries
$d_{jj}$, along the diagonal, such that
$\Phi_D: \M_{2^n}\rightarrow M_{2^n}$ defined by
\begin{equation}\label{MQD}
\Phi_D\left(\frac{1}{2^n}\left(I + c \vec{r}\cdot\vec{\lambda}\right)\right) = \frac{1}{2^n}\left(I
+ c D\vec{r}\cdot \vec{\lambda}\right)
\end{equation}
is a quantum operator.  Just as in the single-qubit setting, (\ref{MQD}) together with the
linearity of $\Phi_D$ yields
$$
\Phi_D(I) = I \ \ \text{and}\ \ \Phi_D(\lambda_j) = d_{jj} \lambda_j\ \ \text{for}\ \ j = 1,
\ldots, 2^{2n} -1.
$$
Because $\Phi_D$ preserves Hermiticity, we know that it has an operator-sum decomposition;
moreover, we know that $\Phi_D$ has $\{\lambda_j\}_{j=0}^{2^{2n}-1}$ as eigenvectors. 
The one-qubit situation, analyzed in the preceding section, suggests that $\Phi_D$ has the form
\begin{equation}\label{PhiDForm}
\Phi_D(\rho) = \sum_{i=0}^{2^{2n}-1}\beta_i \lambda_i \rho\lambda_i
\end{equation}
for some real constants $\{\beta_j\}_{j=1}^{2^{2n}-1}$. 

Using (\ref{PPP}), it is easy to check that
\begin{equation}\label{MQS}
\lambda_i\lambda_j \lambda_i = \pm \frac{1}{2^{n-1}}\lambda_j \ \ \text{for}\ \  i,j \in \{0,1, 2, \ldots, 2^{2n} -1\}
\end{equation}
so that $\Phi_D$, given by (\ref{PhiDForm}), will have the right eigenvectors.  We need only arrange for $\Phi_D$ to
have the correct eigenvalues (namely, $1$, $d_{1,1}$, \ldots, $d_{2^{2n}-1,2^{2n}-1}$).

 We need a way to keep track of the signs that appear on the right of (\ref{MQS}). Let $C$
be the $4\times 4$ matrix of $1$'s and $-1$'s defined by
$$
\sigma_i\sigma_j \sigma_i = c_{ij}\sigma_j
$$
so that $C$ is the matrix of coefficients of the system (\ref{OQS}) of the preceding section. 
Note that $C$ has $1$'s along its main diagonal and $C/2$ is a symmetric, orthogonal matrix.  Moreover the
``one-qubit'' Algoet-Fujiwara conditions---$\beta_j\ge 0$ for $j=0,1,2,3$ where the $\beta_j$'s
are given by (\ref{Bvalues})---are equivalent to the requirement that the column vector 
$$
\frac{1}{4}C\left[\begin{array}{c}1\\d_{11}\\d_{22}\\d_{33}\end{array}\right]
$$
have nonnegative components.  

  Now let $F$ be the $16\times 16$ ``sign'' matrix corresponding to (\ref{MQS}) in the $n=2$ qubit
situation:  $\lambda_i\lambda_j \lambda_i = f_{ij}\lambda_j/2$.  It's not difficult to see that $F= C\otimes C$; for
example, to find the ``upper-left'' $4\times 4$ block of $F$, one calculates
\begin{align}
(\sigma_0 \otimes \sigma_j)(\sigma_0\otimes \sigma_k)(\sigma_0 \otimes \sigma_j) =
(\sigma_0\sigma_0\sigma_0)\otimes(\sigma_j\sigma_k\sigma_j) = c_{00}c_{jk}\sigma_0\otimes \sigma_k
\end{align}
so that $c_{00}C$ is the upper-left block of $F$, which is the appropriate submatrix in the Kronecker
product.   Thus when $n=2$, the matrix of coefficients of the $\beta_i$'s in the $16\times 16$ system that results when
$\lambda_0$ throught
$\lambda_{15}$ are subsituted into (\ref{PhiDForm}) is $\frac{1}{2}C\otimes C$, and thus, because the inverse of
$\frac{1}{2}C\otimes C$ is $\frac{1}{8}C\otimes C$,  the necessary and sufficient conditions for
$\Phi_D$ to be completely positive is that
$$
\frac{1}{8}(C\otimes C) \left[\begin{array}{c}1\\d_{1,1}\\\vdots
\\d_{15,15}\end{array}\right]
$$
have nonnegative components.  Moreover, these components are precisely the $\beta_i$'s in (\ref{PhiDForm}) that lead
to an operator-sum decomposition of $\Phi_D$.

In complete generality, we have the following.   
\begin{thm} [Algoet-Fujiwara Conditions for $n$-qubits]\label{TBD} The diagonal linear superoperator
$\Phi_D:\M_{2^n}\rightarrow \M_{2^n}$ defined by $($\ref{MQD}$)$ is completely positive if and only if the column
vector
$$
\frac{1}{2^{n+1}}\left[\begin{array}{rrrr}1&1&1&1\\1&1&-1&-1\\1&-1&1&-1\\1&-1&-1&1\end{array}\right]^{\otimes\, n}\ \
\left[\begin{array}{c}1\\d_{1,1}\\\vdots
\\d_{2^{2n}-1,2^{2n}-1}\end{array}\right]
$$
has nonnegative components. Moreover, if $\beta_{j-1}$ denotes $j$-th component of this column vector $($for $j=1$,
\ldots, $2^{2n}$ $)$, then $\Phi_D$ has operator-sum decomposition
$$
\Phi_D(\rho) = \sum_{j=0}^{2^{2n}-1}\beta_j \lambda_j\rho\lambda_j.
$$
\end{thm}

\begin{acknowledgments}
The authors would like to thank Professor Alan McRae for several informative
conservations. They also wish to thank the referee for providing important references as well as suggestions that
improved the exposition.
\end{acknowledgments}   

\end{document}